\documentstyle[12pt]{article}

\begin{document}

\setcounter{page}{1}

\pagestyle{plain} \vspace{1cm}
\begin{center}
\Large{\bf Observational constraints on Chaplygin cosmology in
 a braneworld scenario with induced gravity and curvature effect}\\
\small \vspace{1cm}  {\bf Kourosh
Nozari$^{a,b,}$\footnote{knozari@umz.ac.ir}},
 \quad {\bf T. Azizi$^{a,}$\footnote{t.azizi@umz.ac.ir}}\quad and \quad {\bf N. Alipour$^{a,}$\footnote{n.alipour@umz.ac.ir}}\\
\vspace{0.5cm} {\it $^{a}$Department of Physics, Faculty of Basic
Sciences, University of Mazandaran,\\
P. O. Box 47416-95447, Babolsar, IRAN\\
\vspace{0.25cm} $^{b}$Research Institute for Astronomy and
Astrophysics of Maragha,\\
P. O. Box 55134-441, Maragha, IRAN}\\
\end{center}
\vspace{1.5cm}
\begin{abstract}
We study cosmological dynamics and late-time evolution of an
extended induced gravity braneworld scenario. In this scenario,
curvature effects are taken into account via the Gauss-Bonnet term
in the bulk action and there is also a Chaplygin gas component on
the brane. We show that this model mimics an effective phantom
behavior in a relatively wider range of redshifts than previously
formulated models. It also provides a natural framework for smooth
crossing of the phantom-divide line due to presence of the Chaplygin
gas component on the brane. We confront the model with observational
data from type Ia Supernovae, Cosmic Microwave Background and Baryon
Acoustic Oscillations to constraint the model parameters space.
\end{abstract}
\newpage
\section{Introduction}
The outcome of supernovae redshift-luminosity distance and also
other observational probes, that the universe is undergoing an
accelerated phase of expansion, has stimulated a lot of attempt to
explain this unexpected feature. Since dynamics of the universe is
described by the Friedmann equation which follows from the Einstein
field equations in four dimensions, all modifications of the
Friedmann equation ultimately affect the Einstein field equations
too. In this respect, modifications to geometric part of the field
equations imply some sort of alternative geometries, while
modifications to matter sector of the theory involve new forms of
energy densities that have not been observed yet. In fact, the
geometric part of the Einstein's field equations can be modified to
incorporate \emph{dark geometry} as
$G_{\mu\nu}+G^{(Dark)}_{\mu\nu}=8\pi G T^{(M)}_{\mu\nu}$.\, In the
same way and within the second viewpoint, Einstein's field equations
can be written as $G_{\mu\nu}=8\pi G
(T^{(M)}_{\mu\nu}+T^{(Dark)}_{\mu\nu})$\, where $T^{(M)}_{\mu\nu}$
and $T^{(Dark)}_{\mu\nu}$ are energy-momentum tensor of ordinary
matter and dark energy respectively.

Within the first viewpoint, a well-studied model of modified gravity
is the Dvali- Gabadadze-Porrati (DGP) braneworld scenario (G. Dvali,
G. Gabadadze \& M. Porrati 2000; G. Dvali et al. 2000; A. Lue 2006
),\, in which our four-dimensional world is a FRW 3-brane embedded
in a five-dimensional Minkowski bulk. The model is characterized by
a cross-over length scale $r_{c}$ such that gravity is a
four-dimensional theory at scales $r\ll r_{c}$ where matter behaves
as pressureless dust. In the self-accelerating DGP branch gravity
leaks out into the bulk at scales $r\gg r_{c}$ and the cosmology
approaches the behavior of a cosmological constant (C. Deffayet
2001; C. Deffayet, G. Dvali \& G. Gabadadze 2002 ). In the
self-decelerating, normal DGP branch, gravity leaks in from the bulk
at scales $r\gg r_{c}$, leading to a cosmology which is in contrast
to the observed late-time acceleration.

Cosmological dynamics in a braneworld setup that treats Gauss-Bonnet
curvature effect and the DGP induced gravity in a unified manner,
has been studied recently (G. Kofinas, R. Maartens \& E.
Papantonopoulos 2003; R. A. Brown et al. 2005; R. -G. Cai, H. -S.
Zhang \& A. Wang 2005; R. A. Brown 2007; H. Maeda, V. Sahni \& Y.
Shtanov 2007; J. -H. He, B. Wang \& E. Papantonopoulos 2007; K.
Nozari \& B. Fazlpour 2008; K. Nozari \& N. Rashidi 2009 ). This
model which is called GBIG-gravity, is a generalized braneworld
scenario that contains both UV (ultra-violet) and IR (infra-red)
limits in a unified manner: It contains stringy effect via the
Gauss-Bonnet (GB) term in the bulk action as the UV sector of the
theory and Induced Gravity (IG) effect which becomes important in
the IR limit. The cosmological dynamics and possible realization of
the phantom-like behavior in this setup are studied recently (M.
Bouhmadi-Lopez \& P. V. Moniz (2008, 2009); K. Nozari and N. Rashidi
2009; K. Nozari, T. Azizi \& M. R. Setare 2009; K. Nozari \& N.
Aliopur 2009).

Within the second viewpoint and focusing on the matter sector of the
Einstein field equations, a well-studied model introduces into
$T_{\mu\nu}$ a dark energy component called the Chaplygin gas (A.
Dev, J. S. Alcaniz \& D. Jain 2003; L. Amendola et al. 2003; O.
Bertolami et al. 2004; M. Biesiada, W. Godlowski \& M. Szydlowski
2005; X. Zhang, F. Wu \& J. Zhang 2006; H. Zhang \& Z. -H. Zhu 2006;
Heydari-Fard \& H. R. Sepangi 2008; H. Zhang, Z. -H. Zhu \& L. Yang
2009; M. H. Mohseni Sadjadi 2009). This model is similar to the DGP
model in the sense that it is also characterized by a cross-over
length scale below which the gas behaves as pressureless dust, and
above which it approaches the behavior of a cosmological constant.
This length scale is expected to be of the same order of magnitude
as the $r_{c}$ scale of the DGP model (M. Roos 2007). Accelerating
Chaplygin gas combined with the decelerating braneworld DGP model
can produce an overall accelerated expansion of the order of
magnitude seen (M. Roos (2007, 2008a,b); M. Bouhmadi-Lopez \& R.
Lazkoz 2007). However, both the self-accelerating DGP model in flat
space and the standard Chaplygin gas model have problems in fitting
with present supernovae data (T. M. Davis 2007).

Recently, astronomical observations with WMAP$7$ have indicated that
the equation of state parameter of dark energy can be less than $-1$
and even can display a transient behavior (E. Komatsu et al. 2010 ).
A simple way to explain this phenomenon is to consider a
non-canonical phantom dark energy (R. R. Caldwell 2002), that
introduces new theoretical facilities and challenges in this field.
Phantom fields are a sort of scalar fields with negative sign for
the kinetic energy term. Indeed, phantom fields suffer from
instabilities due to violation of the null energy condition, and a
phantom universe eventually ends up with a Big Rip singularity (R.
R. Caldwell, M. Kamionkowski \& N. N. Weinberg). Thus it follows
immediately that there must be some alternative approaches to
realize a phantom-like behavior without introducing any phantom
field in the model. With phantom-like behavior, we mean the growth
of the effective dark energy density with cosmic time and in the
same time, the effective equation of state parameter should stay
always less than $-1$. In this regard, it has been shown that the
normal, non-self-accelerating branch of the DGP scenario has the
potential to explain the phantom-like behavior without introducing
any phantom fields on the brane (V. Sahni \& Y. Shtanov 2003; A. Lue
\& G. D. Starkman 2004; R. Maartens \& E. Majerotto 2006). The main
problem associated to this model and its extended versions (L. P.
Chimento, R. Lazkoz, R. Maartens 2006; M. Bouhmadi-Lopez \& P. V.
Moniz 2008) is that the effective phantom picture always breaks down
at some point as the effective dark energy density will cease to be
positive at a redshift in the past.

With these preliminaries, in this paper we study cosmological
dynamics and late-time evolution of an extended induced gravity
braneworld scenario. In this scenario, curvature effects are taken
into account via the Gauss-Bonnet term in the bulk action and there
is also a Chaplygin gas component on the brane. Our motivation for
incorporation of the GB correction in the DGP setup is that the DGP
model gives just the infra-red modification of the general
relativity. It would be expected that a consistent DGP braneworld
model would have also ultraviolet modifications as well, associated
to high-energy stringy effects at earlier times. This expectation
can be fulfilled via incorporation of the GB term in the bulk
action. By including a Chaplygin gas fluid on the brane, there will
be no breakdown of the phantom-like prescription in this model. The
presence of the Chaplygin gas component also provides a smooth
crossing of the cosmological constant line in this scenario.

The paper is organized as follows: In section $2$ we start from the
action of the scenario and derive the corresponding Friedmann
equation. Then, we study cosmological dynamics of this generalized
braneworld setup and investigate possible realization of the phantom
mimicry without introducing a phantom field. In section $3$, we
investigate possible crossing of the phantom divide line by the
effective equation of state parameter and also we study dynamical
screening of the brane cosmological constant. We constraint the
model parameter space by using the observational data from type Ia
Supernovae, Cosmic Microwave Background and Baryon Acoustic
Oscillations data in section $4$. Finally, our summery and
conclusions are presented in section $5$.

\section{The setup}
We start with the following action
$$S=\frac{1}{2\kappa_{5}^{2}}\int d^{5}x\sqrt{-g^{(5)}}\bigg\{
R^{(5)}-2\Lambda_{5}+\beta\Big[(R^{(5)})^{2}-4R_{ab}^{(5)}R^{(5)ab}+R_{abcd}^{(5)}R^{(5)abcd}\Big]\bigg\}$$
\begin{equation}
+\frac{r_{c}}{2\kappa_{5}^{2}}\int_{y=0} d^{4}x\sqrt{-g}\bigg( R
-\frac{\kappa_{5}^{2}}{r_{c}}\lambda\bigg)+ {\cal{L}}_{m}\,,
\end{equation}
where the first term shows the usual Einstein-Hilbert action in $5$D
bulk with the Gauss-Bonnet term  and $\beta$ is the Gauss-Bonnet
coupling constant. The assumption that $\beta$ is non-negative is
motivated by string theory, where typically $\beta\propto
\ell^{2}_{string}$ (G. Kofinas, R. Maartens \& E. Papantonopoulos
2003; R. A. Brown 2007 ).\, $r_{c}$ is the DGP crossover scale,
$\lambda$ is tension of the brane, and ${\cal{L}}_{m}$ is the matter
field Lagrangian on the brane. In our case, ${\cal{L}}_{m}$
containes ordinary matter (CDM) and a generalized Chaplygin gas
component with equation of state of the form
$p_{ch}=-\frac{A}{\rho_{ch}^{\alpha}}$. The cosmological dynamics on
the brane obeys the following generalized Friedmann equation (M.
Bouhmadi-Lopez \& P. V. Moniz 2008, 2009; K. Nozari and N. Rashidi
2009 )
\begin{eqnarray}
 \bigg[1+\frac{8}{3}\beta\Big(H^{2}+\frac{\Phi}{2}+\frac{K}{a^{2}}\Big)\bigg]^{2}
 \Big(H^{2}-\Phi+\frac{K}{a^{2}}\Big)=\bigg[r_{c}\Big(H^{2}+\frac{K}{a^{2}}\Big)-
 \frac{\kappa^{2}_{5}}{6}\Big(\rho+\rho_{ch}+\lambda\Big)\bigg]^{2}.
 \end{eqnarray}
The bulk contains a black hole mass and a cosmological constant, so
that $\Phi$ is defined as
$\Phi+2\beta\Phi^{2}=\frac{\Lambda_{5}}{6}+\frac{\Upsilon}{a^{4}}$.
The bulk black hole mass originates on the bulk Weyl tensor so that
the bulk reduces to Schwarzschild-AdS$_{5}$ if $\beta=0$ and to
AdS$_{5}$ if $\beta=0=\Upsilon$. In which follows, we restrict
ourselves to the case that the bulk black hole mass vanishes,
$\Upsilon=0$\,, and therefore
$\Phi+2\beta\Phi^{2}=\frac{\Lambda_{5}}{6}$. In this case the bulk
cosmological constant is given by
$\Lambda_{5}=\frac{-6}{l^{2}}+\frac{12\beta}{l^{4}}$, where $l$ is
the bulk curvature. Assuming $\Lambda_{5}=0$ corresponding to a
Minkowski bulk, for a spatially flat FRW brane ($K=0$), the
Friedmann equation would be as follows
\begin{equation}
\bigg[1+\frac{8}{3}\beta\Big(H^{2}+\frac{\Phi}{2}\Big)\bigg]^{2}(H^{2}-\Phi)=\bigg[r_{c}H^{2}
-\frac{\kappa^{2}_{5}}{6}\Big(\rho+\rho_{ch}+\lambda\Big)\bigg]^{2}.
\end{equation}
With $\Lambda_{5}=0$\,, we find $\Phi=0$ or
$\Phi=-\frac{1}{2\beta}$. Our forthcoming arguments will be based on
the choice $\Phi=0$. Now we look at the conservation equation. In
our setup, there is no energy exchange between bulk and brane.
Therefore, total matter/energy budget on the brane is conserved by
virtue of the Bianchi's identity so that
$\dot{\rho}_{tot}+3H(\rho_{tot}+p_{tot})=0$, where $p_{tot}$\, and
$\rho_{tot}$\, are total pressure and energy density on the brane
respectively. Restricting further so that conservation holds for two
matter components separately, we have for the CDM component
\begin{equation}
\dot{\rho}_{m}+3H\rho_{m}=0
\end{equation}
which integrates to the usual $\rho_{m}=\rho_{m_{0}}(1+z)^3$,
whereas for the Chaplygin gas component, the continuity equation can
be written as
\begin{equation}
\dot{\rho}_{ch}+3H(\rho_{ch}+p_{ch})=0.
\end{equation}
Since $p_{ch}=-\frac{A}{\rho_{ch}^{\alpha}}$\, (M. Roos 2007; M.
Bouhmadi-Lopez \& R. Lazkoz 2007 ), we find
\begin{equation}
\rho_{ch}=(\rho_{ch})_{0}\Big[A_{s}+\frac{1-A_{s}}{a^{3(1+
\alpha)}}\Big]^{\frac{1}{1+\alpha}}\,,
\end{equation}
where $A_{s}=\frac{A}{(\rho_{ch})_{0}^{1+\alpha}}$. Using the
definition of the redshift parameter $z$ as
$\frac{a}{a_{0}}=\frac{1}{1+z}$, and setting $a_{0} = 1$ for
convenience, we find
\begin{equation}
\rho_{ch}=(\rho_{ch})_{0}\Big[A_{s}+(1-A_{s})(1+z)^{3(\alpha+1)}\Big]^{\frac{1}{1+\alpha}}.
\end{equation}
This expression supports the interest on Chaplygin cosmologies since
it reflects the fact that the energy density of such fluids
interpolates between dust and a cosmological constant (A. Y.
Kamenshchik, U. Moschella \& V. Pasquier 2001; N. Bilic, G. B.
Tupper \& R. D. Viollier 2002; M. C. Bento, O. Bertolami \& A. A.
Sen 2002). After these preliminaries, we are looking for the
late-time behavior of the normal branch of this chaplygin GBIG
scenario. Defining the cosmological parameters as\,
$\Omega_{m}=\frac{\kappa^{2}_{4}\rho_{m_{0}}}{3H^{2}_{0}}$\,,
$\Omega_{\Lambda}=\frac{\kappa^{2}_{4}\Lambda}{3H^{2}_{0}}$\,,
$\Omega_{r_{c}}=\frac{1}{4r_{c}^{2}H^{2}_{0}}$\,\,,
$\Omega_{\beta}=\frac{8}{3}\beta H^{2}_{0}$\, and \,
$\Omega_{ch}=\frac{(\rho_{ch})_{0}\kappa^{2}_{4}}{3H^{2}_{0}}$,\,
the Friedmann equation on the brane, equation (3), can be expressed
in a dimensionless form as follows
\begin{eqnarray}
E^{2}(z)=-2\sqrt{\Omega_{r_{c}}}E(z)\big[1+\Omega_{\beta}E^{2}
(z)\big]+\Omega_{m}(1+z)^{3}+\Omega_{ch}[A_{s}+(1-A_{s})
(1+z)^{3(\alpha+1)}]^{\frac{1}{1+\alpha}}+\Omega_{\Lambda}\,,
\end{eqnarray}
where $E(z)=\frac{H}{H_{0}}$. We assume $0<A_{s}<1$ and
$1+\alpha>0$. With these conditions, it is possible to realize a de
Sitter phase at late time (M. Bouhmadi-Lopez \& R. Lazkoz 2007 ). As
an important ingredient of the model, the following constraint
equation can be obtained from (8) by setting $z=0$
\begin{eqnarray}
1+2\sqrt{\Omega_{r_{c}}}(1+\Omega_{\beta})=\Omega_{m}+\Omega_{ch}+\Omega_{\Lambda}.
\end{eqnarray}
Note that based on this relation, the region
$\Omega_{m}+\Omega_{ch}+\Omega_{\Lambda}<1$ in the model parameters
space is physically unacceptable. Taking the time derivative of
equation (9), and using the continuity equation for matter on the
brane, the Hubble rate can be deduced as follows
\begin{equation}
\frac{\dot{H}}{H_{0}^{2}}=-\frac{3}{2}\frac{[E(z)(1+z)^{3}]\Big[\Omega_{m}+
\frac{3(1-A_{s})\Omega_{ch}(1+z)^{3\alpha}}{[A_{s}+(1-A_{s})(1+z)^{3\alpha}]^{\frac{\alpha}{\alpha+1}}}\Big]}
{E(z)+\sqrt{\Omega_{r_{c}}}[1+\Omega_{\beta}E^{2}(z)]+2\sqrt{\Omega_{r_{c}}}\Omega_{\beta}E^{2}(z)}\,.
\end{equation}
With the previous constraints on $A_s$, and assuming that
$\beta\ge0$ ( motivated by string theory as $\beta\propto
\ell^{2}_{string}$), it is obvious that this relation always has a
negative value in the whole physically admissible parameters space.
This feature is plotted in fig $1$\footnote{ The numerical values of
cosmological parameters used to plot figures of this paper are taken
from Table 2.}. Since $\dot{H}<0$, the Hubble parameter decreases as
the brane expands, consequently there is no super-acceleration in
this braneworld universe. Therefore, the brane does not hit a big
rip singularity as its fate. Note also that $\dot{H}$ vanishes when
$z\rightarrow-1$, while $H$ is positive at this limit. This is a
reflection of late-time de Sitter character of the model. It has
been shown that it is possible to have \emph{Big Freeze} singularity
in FRW universe filled with a generalized Chaplygin gas (A. V. Yurov
et al. 2008; M. Bouhmadi-Lopez, P. F. Gonzalez-Diaz \& P.
Martin-Moruno 2008; M. Bouhmadi-Lopez et al. (2009,2010) ). So,
essentially appearance of this type of singularity in our framework
is probable too. Nevertheless, existence of induced gravity on the
brane and the Gauss-Bonnet term in the bulk action may help to
overcome this difficulty. This issue needs further investigations
and we are going to study its separately.
\begin{figure}[htp]
\begin{center}
\includegraphics{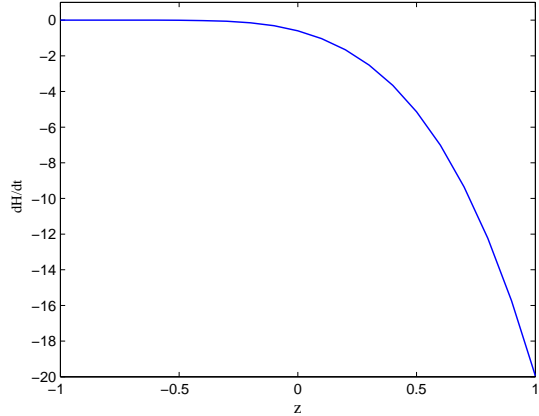}
\end{center}
\vspace{5.5 cm} \caption{\small { Variation of
$\frac{\dot{H}}{H^{2}_{0}}$ with redshift. The parameter values to
plot this figure are taken from Table 2.}}
\end{figure}
The deceleration parameter, $q$ depends on $\dot{H}$ through the
relation $q=-[\frac{\dot{H}}{H^{2}}+1]$, which can be expressed in
the following form
\begin{equation}
q=-\bigg(1-\frac{3}{2}\frac{[\frac{(1+z)^{3}}{E(z)}]\Big(\Omega_{m}+
\frac{3(1-A_{s})\Omega_{ch}(1+z)^{3\alpha}}
{[A_{s}+(1-A_{s})(1+z)^{3\alpha}]^{\frac{\alpha}{\alpha+1}}}\Big)}
{[E(z)+\sqrt{\Omega_{r_{c}}}
(1+\Omega_{\beta}E^{2}(z))+2\sqrt{\Omega_{r_{c}}}
\Omega_{\beta}E^{2}(z)]}\bigg)\,.
\end{equation}
In figure $2$, the dimensionless deceleration parameter $q$ is
plotted versus the redshift for a fixed set of the parameters as are
presented in table 2. In this model, the universe enters the
accelerating phase at $z\approx 0.13$.
\begin{figure}[htp]
\begin{center}
\includegraphics{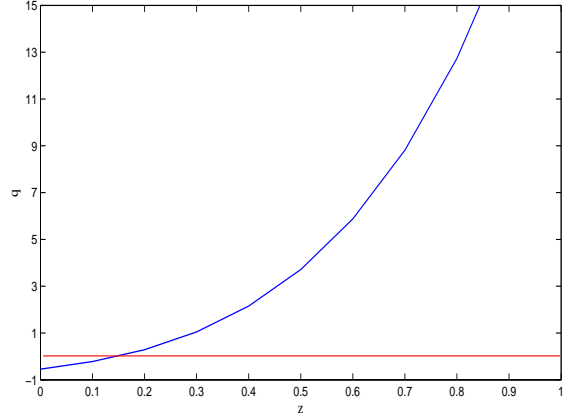}
\end{center}
\vspace{4.5cm} \caption{\small {Variation of $q$ versus the
redshift. Parameters values are taken from Table 2.}}
\end{figure}

\section{Crossing the phantom divide}
In this section, we show that a phantom-like behavior can be
realized on the brane, in a relatively wider range of redshifts than
previously formulated models ( see for instance R. Maartens \& E.
Majerotto 2006; L. P. Chimento, R. Lazkoz, R. Maartens 2006; M.
Bouhmadi-Lopez \& P. V. Moniz 2008). This phantom-like behavior
occurs without including any phantom matter. The phantom-like
prescription is based on the definition of an effective energy
density which is corresponding to a balance between the cosmological
constant and geometrical effects encoded in the Hubble rate
evolution (V. Sahni \& Y. Shtanov 2003; A. Lue \& G. D. Starkman
2004; V. Sahni 2005). This behavior is based on the definition of an
effective energy density $\rho_{eff}$, which increases with cosmic
time and an effective equation of state parameter always less than
$-1$. More precisely, the effective description is inspired in
writing down the modified Friedmann equation of the brane as the
usual $4$D Friedmann equation(A. Lue \& G. D. Starkman 2004; M.
Bouhmadi-Lopez \& P. V. Moniz 2008 ), so that
\begin{eqnarray}
H^{2}=\frac{\kappa^{2}_{4}}{3}(\rho_{m}+\rho_{eff}).
\end{eqnarray}
Using equations (8) and (12)\,, we find
\begin{equation}
\rho_{eff}=\frac{3H_{0}^{2}}{\kappa_{4}^{2}}\Big[-2\sqrt{\Omega_{r_{c}}}E(z)\big[1+\Omega_{\beta}E^{2}
(z)\big]+\Omega_{ch}[A_{s}+(1-A_{s})(1+z)^{3(\alpha+1)}]^{\frac{1}{1+\alpha}}+\Omega_{\Lambda}\Big]
\end{equation}
By definition, phantom-like prescription breaks down if
$\rho_{eff}\leq0$. Figure $3$ shows variation of $\rho_{eff}^{(DE)}$
versus $\alpha$ and redshift in this model. The phantom-like
behavior can be realized for all values of $\alpha$ that
$1+\alpha>0$.
\begin{figure}[htp]
\begin{center}\includegraphics{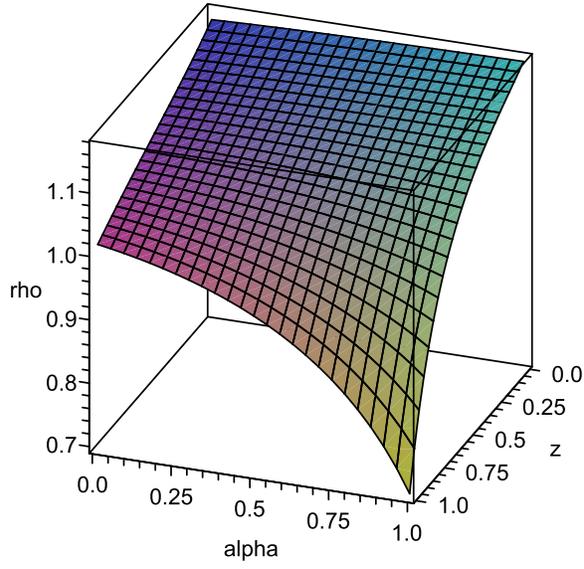}
 \vspace{7cm}
\end{center}
 \caption{\small {Variation of the effective dark energy versus
$\alpha$ and redshift. Parameters values are taken from Table 2.}}
\end{figure}
Figure $4$ shows variation of $\rho_{eff}^{(DE)}$ versus $A_{s}$ and
redshift in this  model. The phantom-like behavior can be realized
if $0<A_{s}<1$.
\begin{figure}[htp]
\begin{center}\includegraphics{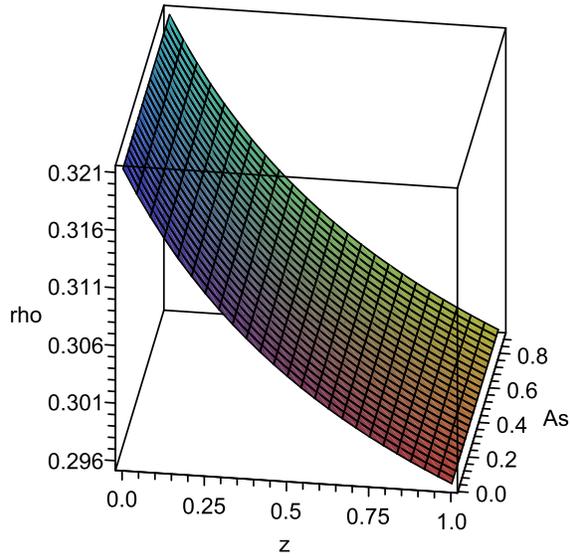}
 \vspace{5cm}
\end{center}
 \caption{\small {Variation of the effective dark energy versus
$A_{s}$ and redshift. Parameters values are taken from Table 2.}}
\end{figure}
Based on this analysis, $\rho_{eff}$ is always positive and grows by
cosmic expansion, and this is a typical phantom-like behavior. Note
that in the absence of the Chaplygin gas component on the brane, the
effective phantom picture breaks down at a redshift in the past
where the effective dark energy density becomes negative. The effect
of the Chaplygin component as a new ingredient added to the GBIG
braneworld scenario, is that the effective phantom-like picture has
no break down in this case. The main point here is the fact that it
is possible essentially to save phantom-like prescription for a wide
range of redshifts in this case. Now, the effective equation of
state parameter $\omega_{eff}$, can be defined using the
conservation equation of the effective energy density on the brane
\begin{eqnarray}
\dot{\rho}_{eff}+3H(1+\omega_{eff})\rho_{eff}=0.
\end{eqnarray}
A straightforward calculation shows (M. Roos 2007; M. Bouhmadi-Lopez
\& R. Lazkoz 2007 )
\begin{equation}
\dot{\rho}_{ch}=-\frac{9HH_{0}^{2}\Omega_{ch}(1-A_{s})
(1+z)^{3(1+\alpha)}}{[A_{s}+(1-A_{s})(1+z)^{3(1+\alpha)}]^{\frac{\alpha}{\alpha+1}}}\,.
\end{equation}
Then the effective energy density evolves as follows

$$\dot{\rho}_{eff}=\frac{-3H_{0}^{2}}{\kappa_{4}^{2}}\bigg\{\frac{3H(1+z)^{3}(1-A_{s})
\Omega_{ch}(1+z)^{3\alpha}}
{[A_{s}+(1-A_{s})(1+z)^{3\alpha}]^{\frac{\alpha}{\alpha+1}}}+$$
\begin{equation}
2\sqrt{\Omega_{r_{c}}}E(z)\big[1+\Omega_{\beta}E^{2}
(z)]\dot{H}H_{0}+2\sqrt{\Omega_{r_{c}}}\dot{H}H_{0}\Omega_{\beta}E^{2}(z)\bigg\}
\end{equation}
Using equations (13) and (16), we find
\begin{equation}
1+\omega_{eff}=\frac{\frac{3H(1+z)^{3}(1-A_{s})\Omega_{ch}(1+z)^{3\alpha}}
{[A_{s}+(1-A_{s})(1+z)^{3\alpha}]^{\frac{\alpha}{\alpha+1}}}+2\sqrt{\Omega_{r_{c}}}E(z)\big[1+\Omega_{\beta}E^{2}
(z)]\dot{H}H_{0}+2\sqrt{\Omega_{r_{c}}}\dot{H}H_{0}\Omega_{\beta}E^{2}(z)}{3E(z)
\Big[-2\sqrt{\Omega_{r_{c}}}E(z)\big[1+\Omega_{\alpha}E^{2}
(z)\big]+\Omega_{ch}[A_{s}+(1-A_{s})(1+z)^{3(\alpha+1)}]^{\frac{1}{1+\alpha}}+\Omega_{\Lambda}\Big]}
\end{equation}
Figure $5$ shows the plot of $1+\omega_{eff}$ versus the redshift.
The universe enters to the phantom phase smoothly at $z\approx0.12$.
It is important to note that this setup realizes a smooth transition
to the phantom phase, the so called \emph{phantom-divide line
crossing}. We note that this smooth crossing behavior cannot be
realized in the absence of the Chaplygin fluid component. Indeed,
without this term, the effective equation of state parameter blows
up in the past and as mentioned previously, the effective phantom
description breaks down.

\begin{figure}[htp]
\begin{center}
\includegraphics{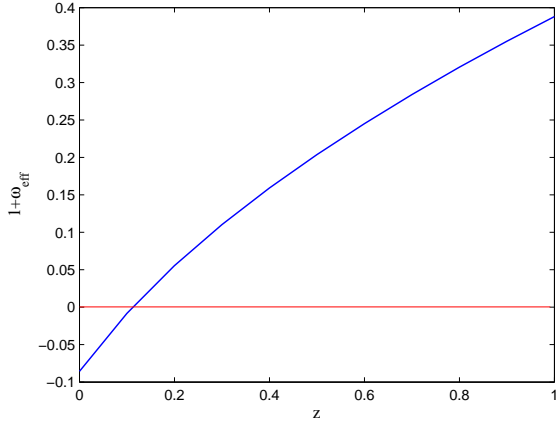}
\end{center}
\vspace{5.5 cm} \caption{\small { $1+w_{eff}$ versus the redshift.
This model realizes (as a result of the existence of the Chpalygin
component on the brane), a smooth crossing of the phantom divide
line. Parameters values are taken from Table 2. }}
\end{figure}

Now we study the phantom-like behavior in another perspective:
\emph{dynamical screening of the brane cosmological constant} (A.
Lue \& G. D. Starkman 2004 ). The normal branch of this model can be
described by the following Friedmann equation
\begin{equation}
H^{2}=\frac{8\pi
G}{3}(\rho_{m}+\rho_{ch})+\frac{\Lambda}{3}-\frac{H}{r_{c}}(1+\frac{8}{3}\beta
H^{2})
\end{equation}
where $\Lambda$ is the brane cosmological constant. We rewrite this
equation as follows
\begin{equation}
H^{2}=\frac{8\pi G}{3}(\rho_{m}+\rho_{ch})+\frac{8\pi
G}{3}\Lambda^{(eff)},
\end{equation}
where by definition
\begin{equation}
\Lambda^{(eff)}=\frac{\Lambda}{3}-\frac{H(1+\frac{8}{3}\beta
H^{2})}{r_{c}}\,.
\end{equation}
This equation means that the brane is extrinsically curved so that
shortcuts through the bulk allow gravity to screen the effects of
the brane cosmological constant at Hubble parameters $H\sim
r_{c}^{-1}$ where $r_{c}$ is the DGP crossover distance. As a
distinctive feature of this model, curvature effect via the
Gauss-Bonnet term and presence of the Chaplygin component on the
brane contribute in the dynamical screening of the brane
cosmological constant. While the role played by the GB term is
obvious, the role of the Chaplygin matter is hidden in the
definition of the Hubble parameter. This feature leads to a
considerable difference relative to the pure DGP case studied in (A.
Lue \& G. D. Starkman 2004 ). Figure $6$ shows the difference
between two scenarios in this respect for some values of
$\Omega_{\beta}$.
\begin{figure}[htp]
 \begin{center}
 \includegraphics{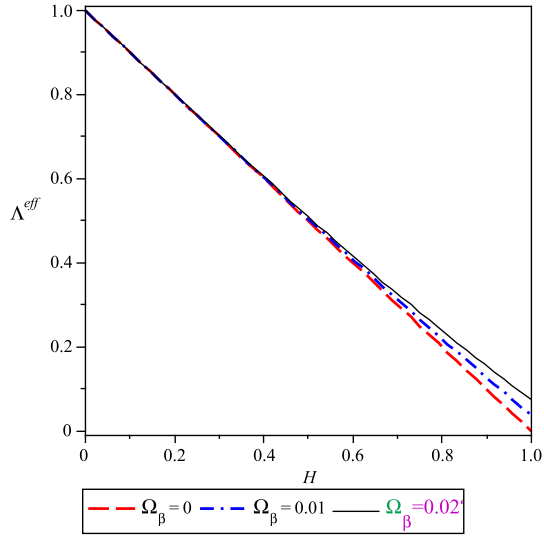}
 \end{center}
 \vspace{7 cm} \caption{\small { $\Lambda^{(eff)}$ versus the Hubble
 parameter for different values of $\Omega_{\beta}$. }}
 \end{figure}
Alternatively, we can define also
\begin{equation}
\rho_{DE}^{(eff)}=\frac{\Lambda}{3}-\frac{H(1+\frac{8}{3}\beta
H^{2})}{r_{c}}.
\end{equation}
Figure $7$ shows variation of $\rho_{DE}^{(eff)}$ versus $z$ in our
proposed setup. It is always positive and grows with cosmic
expansion; a typical phantom-like behavior.
\begin{figure}[htp]
 \begin{center}
 \includegraphics{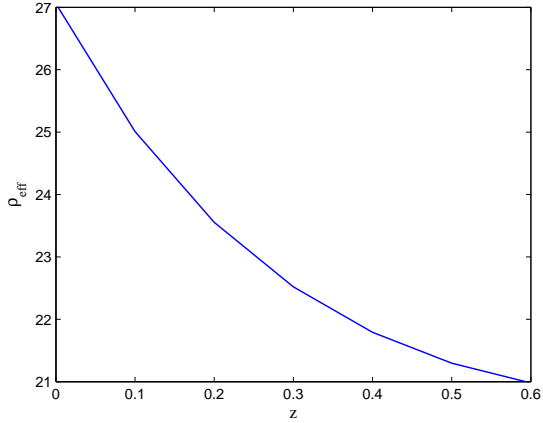}
 \end{center}
 \vspace{6 cm} \caption{\small { $\rho_{DE}^{(eff)}$ versus the redshift.
 Parameters values are taken from Table 2. }}
 \end{figure}

\section{Observational Constraints}
In this section we study the constraints imposed on the GBIG
chaplygin model parameters using the observational data such as the
gold sample of SNIa combined with the information from the BAO
measurement by SDSS and the CMB shift parameter from WMAP7
observations.\\

\textbf{A: SNIa data}\\

In this section we use the 156 new gold sample supernovae Ia data
compiled in (A. G. Riess et al. 2007) to fit the model. This
observation directly measures the apparent magnitude $m$ of a
supernovae and its redshift $z$. The apparent magnitude $m$ is
related to the luminosity distance $d_{L}$ of the supernovae through
\begin{equation}
\mu=m-M=\log_{10}d_{L}+5\log_{10}\big(\frac{cH_{0}}{Mpc}\big)+25,
\end{equation}
where $M$ is the absolute magnitude that is believed to be constant
for all Type Ia supernovae and $d_{L}$ is the luminosity distance
that in a flat universe can be expressed as follows
\begin{equation}
d_{L}=(1+z)\int_{0}^{z}\frac{dz'}{E(z';\theta)}
\end{equation}
where $\theta$ denotes the model parameters. From equation (9), the
$E(z)$ is given by
\begin{equation}
E(z)=\frac{1}{{\Omega_{\beta}\sqrt\Omega_{{r}_{c}}}}\Big[P^{\frac{1}{3}}+
\frac{1-12\Omega_{\beta}\Omega_{{r}_{c}}}{P^{\frac{1}{3}}}-1\Big]
\end{equation}
where
$$
P\equiv18\Omega_{\beta}\Omega_{{r}_{c}}+54Q\Omega_{\beta}^{2}\Omega_{{r}_{c}}
-1+6\sqrt{3\Omega_{\beta}\Omega_{{r}_{c}}}\sqrt{16\Omega_{\beta}
\Omega_{{r}_{c}}^2+18\Omega_{\beta}\Omega_{{r}_{c}}Q
+24\Omega_{{r}_{c}}\Omega_{\beta}^{2}Q^{2}-\Omega_{{r}_{c}}-Q},$$
and
$$Q\equiv\Omega_{m}(1+z)^{3}+\Omega_{ch}[A_{s}+(1-A_{s})
(1+z)^{3(\alpha+1)}]^{\frac{1}{1+\alpha}}+\Omega_{\Lambda}.$$ For
the statistical analysis of the supernovae data we perform the
$\chi^{2}$ statistics for the model parameter that is calculated as
\begin{equation}
\chi^{2}_{SN}(\theta)=\sum_{i=1}^{N}\frac{[\mu_{obs}(z_{i})-\mu_{th}(z_{i})]^{2}}{\sigma_{i}^{2}},
\end{equation}
In this relation, $N=156$ is the number of SNIa data points,
$\mu_{obs}$ is the observed distance modulus and the $\sigma_{i}$ is
the uncertainty in the observed distance modulus, which is assumed
to be Gaussian and uncorrelated so that the likelihood is
proportional to $e^{\frac{-\chi^2}{2}}$. The parameter
$$\bar{M}=5\log_{10}\big(\frac{cH_{0}}{Mpc}\big)+25\,,$$ is a \emph{nuisance}
parameter and should be marginalized (integrated out).  This can be
done by following the techniques described in (M. S. Movahed, M.
Farhang \& S. Rahvar 2009 ) and can be performed by expanding the
$\chi^{2}_{SN}$ of equation (25) with respect to $M$ as
\begin{equation}
\chi^{2}_{SN}(\theta)=\tilde{A}-2M\tilde{B}+M^{2}\tilde{C}
\end{equation}
where
\begin{equation}
\tilde{A}(\theta)=\sum_{i=1}^{N}\frac{[\mu_{obs}(z_{i})-\mu_{th}(z_{i},M=0,\theta)]^{2}}{\sigma_{i}^{2}},
\end{equation}
and
\begin{equation}
\tilde{B}(\theta)=\sum_{i=1}^{N}\frac{\mu_{obs}(z_{i})-\mu_{th}(z_{i},M=0,\theta)}{\sigma_{i}^{2}}\,
,\quad\quad\tilde{C}=\sum_{i=1}^{N}\frac{1}{\sigma_{i}^{2}}.
\end{equation}
Equation (25) has a minimum for $M=\frac{\tilde{B}}{\tilde{C}}$ \,
at\, $
\tilde{\chi}^{2}_{SN}(\theta)=\tilde{A}(\theta)-\frac{\tilde{B}^{2}(\theta)}{\tilde{C}}$.
Using this equation the best fit values of model parameters as the
values that minimize $\chi^{2}_{SNIa}(\theta)$ can be obtained.

For the likelihood analysis we marginalize the likelihood function
$L =\exp(-\chi^{2}/2)$ over $h$ and $\Omega_{m}=\Omega_{b}$ where
$\Omega_{b}$ is the energy density of the barionic matter. We
adopted Gaussian priors such that $h=0.705$ from the WMAP7 (K.
Komatsu et al. 2009 ) and
$\Omega_{b}=0.0456$. Table $1$ summarizes these priors. \\

\begin{table}
\begin{center}
\caption{Priors on the parameter space used in the likelihood
analysis.} \vspace{0.5 cm}
\begin{tabular}{|c|c|c|c|c|c|c|c|c|c|}
  \hline
  \hline Parameter&\,\,\,Prior&\\
  \hline $\Omega_{ch}$&  0.00\, - \, 1.00& Top Hat\\
  \hline $\Omega_{\lambda}$&  0.00\, - \, 1.00 &Top Hat\\
  \hline $\Omega_{\beta}$& 0.00\, -\, 1.00&  Top Hat \\
  \hline $\Omega_{b}h^{2}$& $0.02265\pm 0.00059$&  Top Hat (BBN) \\

   \hline
\end{tabular}
\end{center}
\end{table}

\textbf{B: CMB shift parameter}\\

We use the CMB data from WMAP7 observation that includes the shift
parameter $\cal{R}$ and the redshift of the decoupling epoch
$z_{*}$. The shift parameter $\cal{R}$ is related to the angular
diameter distance to the last scattering surface, the comoving size
of the sound horizon at $z_{*}$ and the angular scale of the first
acoustic peak in the CMB power spectrum of the temperature
fluctuations. The CMB shift parameter is approximated by (see for
more details R. Lazkoz \& E. Majerotto (2007 ). )
\begin{equation}
{\cal{R}}=\sqrt{\Omega_{m}}H_{0}r(z_{*}),
\end{equation}
where $r(z)$ is the comoving distance to the redshift $z$ defined by
\begin{equation}
r(z)=\int_{0}^{z}\frac{1}{H(z)}dz.
\end{equation}
The constraints on a typical model using CMB shift is obtained from
minimization of the quantity
\begin{equation}
\chi^{2}_{CMB}=\frac{[{\cal{R}}_{obs}-{\cal{R}}_{th}]^{2}}{\sigma_{CMB}^{2}},
\end{equation}
where ${\cal{R}}_{obs}$ is the observed value of the CMB shift
parameter performed from WMAP7 observation (K. Komatsu et al 2009)
and ${\cal{R}}_{th}$ is corresponding to the theoretical value
calculated from equation
(29).\\

\textbf{C: BAO observation}\\

The baryonic acoustic oscillation (BAO) peak detected in the SDSS
luminous red Galaxies (LRG) is another tool to test the model
against observational data. BAO are described in terms of a
dimensionless parameter
\begin{equation}
{\cal{A}}=\sqrt{\Omega_{m}}\bigg[\frac{H_{0}d_{L}^{2}(z;\theta)}
{H(z_{sdss};\theta)z_{sdss}^{2}(1+z_{sdss})^{2}}\bigg]^{1/3}\,.
\end{equation}
The best fit values of the model parameters can be determined by
constructing minimization of the quantity
\begin{equation}
\chi^{2}_{SDSS}=\frac{[{\cal{A}}_{obs}-{\cal{A}}_{th}]^{2}}{\sigma_{sdss}^{2}}.
\end{equation}
The observed value ${\cal{A}}_{obs}$ from the LRG is
${\cal{A}}_{obs}= 0.469\Big(\frac{n_{s}}{0.98}\Big)^{-0.35} \pm
0.017$ measured at $z_{sdss}=0.35$ (R. Lazkoz \& E. Majerotto, 2007
). Here $n_{s}=0.963$ is the spectral index as measured by WMAP
seven year observations (K. Komatsu et al 2009 ). It is important to
note that the above observational data are uncorrelated form each
other, since they are given by different experiments and methods.
Then we can construct a joint analysis of them as
\begin{equation}
\chi^{2}_{tot}=\chi^{2}_{SN}+\chi^{2}_{CMB}+\chi^{2}_{SDSS}.
\end{equation}
With these preliminaries, we have obtained the best fit parameters
of the GBIG chaplygin model for SNIa data, the joint analysis of the
SNIa and CMB, and finally the combined analysis of the total
datasets. Table $2$ indicates the results of the observational
constraints on the free parameters of the model. In figure $8$,
using the best fit values of the model parameters and observational
data from gold sample, we have compared the theoretical predictions
of the distance modulus in our model. Figure $9$ shows the
marginalized relative likelihood with respect to parameter $A_S$
fitted with SNIa gold sample, SNIa+CMB and SNIa+CMB+SDSS
experiments. We plot the joint confidence interval of
$\Omega_{{r}_{c}}$ and $A_{S}$ in figure $10$ in $1\sigma$,\,
$2\sigma$ and $3\sigma$ level of confidence for the mentioned
datasets.

Finally and for completeness of discussions, we compare our
Chaplygin GBIG scenario with $\Lambda$DGP and $\Lambda$CDM ( see for
instance (W. J. Percival et al. 2007 ) for a general framework).
Using the data set in the previous subsections, we obtain the best
fit parameters of the $\Lambda$DGP and $\Lambda$CDM models in table
$3$. Comparing tables $2$ and $3$, we see that in our model the best
fit values ( which are given by $\chi^{2}/N_{d.o.f}$) are more
reasonable than $\Lambda$DGP and $\Lambda$CDM ones on observational
ground. In figure $11$ we have plotted the marginalized relative
likelihood with respect to parameter $\Omega_{m}$ fitted with SNIa
gold sample, SNIa+CMB and SNIa+CMB+SDSS experiments. In summary, as
tables $2$ and $3$ show, our Chaplygin GBIG scenario has better
agreement with observations than $\Lambda$DGP and $\Lambda$CDM. In
other words, existence of a Chaplygin component on the DGP brane
brings the scenario to be more viable in observational viewpoint
than the pure $\Lambda$DGP or $\Lambda$CDM. This is a result of
wider parameter space available here which leads to further degrees
of freedom.
\begin{table}
\begin{center}
\caption{The best values for the parameters of GBIG chaplygin model
by fitting with SNIa Gold sample, SNIa+CMB and SNIa+CMB+SDSS
experiments in a flat background.} \vspace{0.5 cm}
\begin{tabular}{|l|c|c|c|c|c|c|c|c|c|}
  \hline
  \hline Observation &$\Omega_{{r}_{c}}\,\,$ & $\Omega_{\beta}$\,\,& $A_S$\,\,
  &$\Omega_{\Lambda}\,\,$&$\Omega_{ch}\,\,$&$\alpha\,\,$&$\chi^{2}_{min}/N_{d.o.f}\,\,$\\
  \hline SNIa( Gold Sample)&0.43& 0.11 &$0.50^{+0.17}_{-0.48}$& 1.45 &0.99&0.99&0.923\\
  \hline SNIa(Gold Sample)+CMB&0.53& 0.31 &$0.12^{+0.15}_{-0.09}$& 1.70 &0.99&-0.98&0.943\\
 \hline SNIa(Gold Sample)+CMB+SDSS&0.51& 0.41 &$0.099^{+0.16}_{-0.08}$& 1.79 &0.99&-0.98&0.992 \\
   \hline
\end{tabular}
\end{center}
\end{table}
\begin{small}
\begin{table}[tbh]
\begin{center}
\caption{The best values for the parameters of $\Lambda$CDM and
$\Lambda$DGP models by fitting with SNIa Gold sample, SNIa+CMB and
SNIa+CMB+SDSS experiments in a flat background.} \vspace{0.5 cm}
\begin{tabular}[tbh]{|l|c|c|c|c|c|c|c|c|c|}
  \hline
  \hline Observation&\,\,$\Lambda$CDM model&\,$\Lambda$DGP model\\
  \hline SNIa&  $\Omega_{m}=0.35^{+0.04}_{-0.04}$, \,$\chi^{2}_{min}/N_{d.o.f}=0.921$&$\Omega_{m}=0.007^{+0.08}_{}$,\,
 $\Omega_{r_{c}}=0.22^{}_{}$,\,$\chi^{2}_{min}/N_{d.o.f}=0.914$\\

  \hline SNIa+CMB&  $\Omega_{m}=0.34^{+0.03}_{-0.04}$, \,$\chi^{2}_{min}/N_{d.o.f}=0.926$

&$\Omega_{m}=0.33^{+0.05}_{-0.12}$,\,
 $\Omega_{r_{c}}=0.00^{}_{}$,\,$\chi^{2}_{min}/N_{d.o.f}=0.926$\\
 \hline SNIa+CMB+SDSS&  $\Omega_{m}=0.32^{+0.04}_{-0.03}$, \,$\chi^{2}_{min}/N_{d.o.f}=0.960$

&$\Omega_{m}=0.34^{+0.02}_{-0.11}$,\,
 $\Omega_{r_{c}}=0.00^{}_{}$,\,$\chi^{2}_{min}/N_{d.o.f}=0.966$\\
\cline{1-2}

   \hline
\end{tabular}
\end{center}
\end{table}
\end{small}

\begin{figure}[htp]
\begin{center}\includegraphics{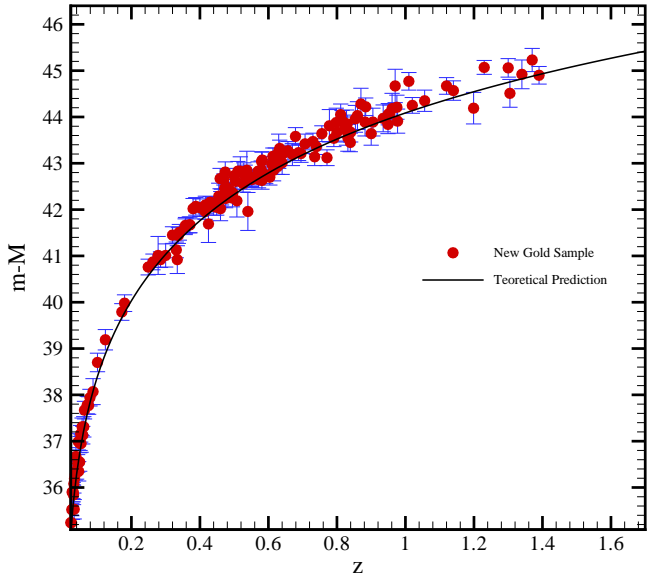} \vspace{6cm}
\end{center}
 \caption{\small {Distance modulus of the SNIa Gold sample versus the redshift. Solid line shows
the best fit values with the corresponding parameters of $h = 0.70$,
$\Omega_{{r}_{c}}=0.43$ and $A_S=0.50$ in $1\sigma$ level of
confidence with $\chi^{2}_{min}=141.368$ for GBIG chaplygin model.}}
\end{figure}

\begin{figure}[htp]
\begin{center}\includegraphics{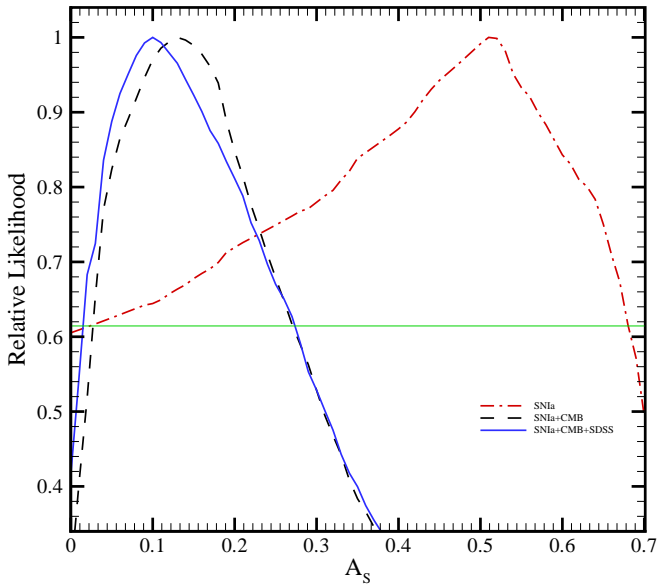} \vspace{6.5cm}
\end{center}
 \caption{\small {Marginalized relative likelihood with respect to parameter $A_S$ fitted with SNIa
gold sample, SNIa+CMB and SNIa+CMB+SDSS experiments. The
intersection of the curve with the horizontal solid line is
corresponding to the bound with $1\sigma$ level of confidence. }}
\end{figure}

\begin{figure}[htp]
\begin{center}
\includegraphics{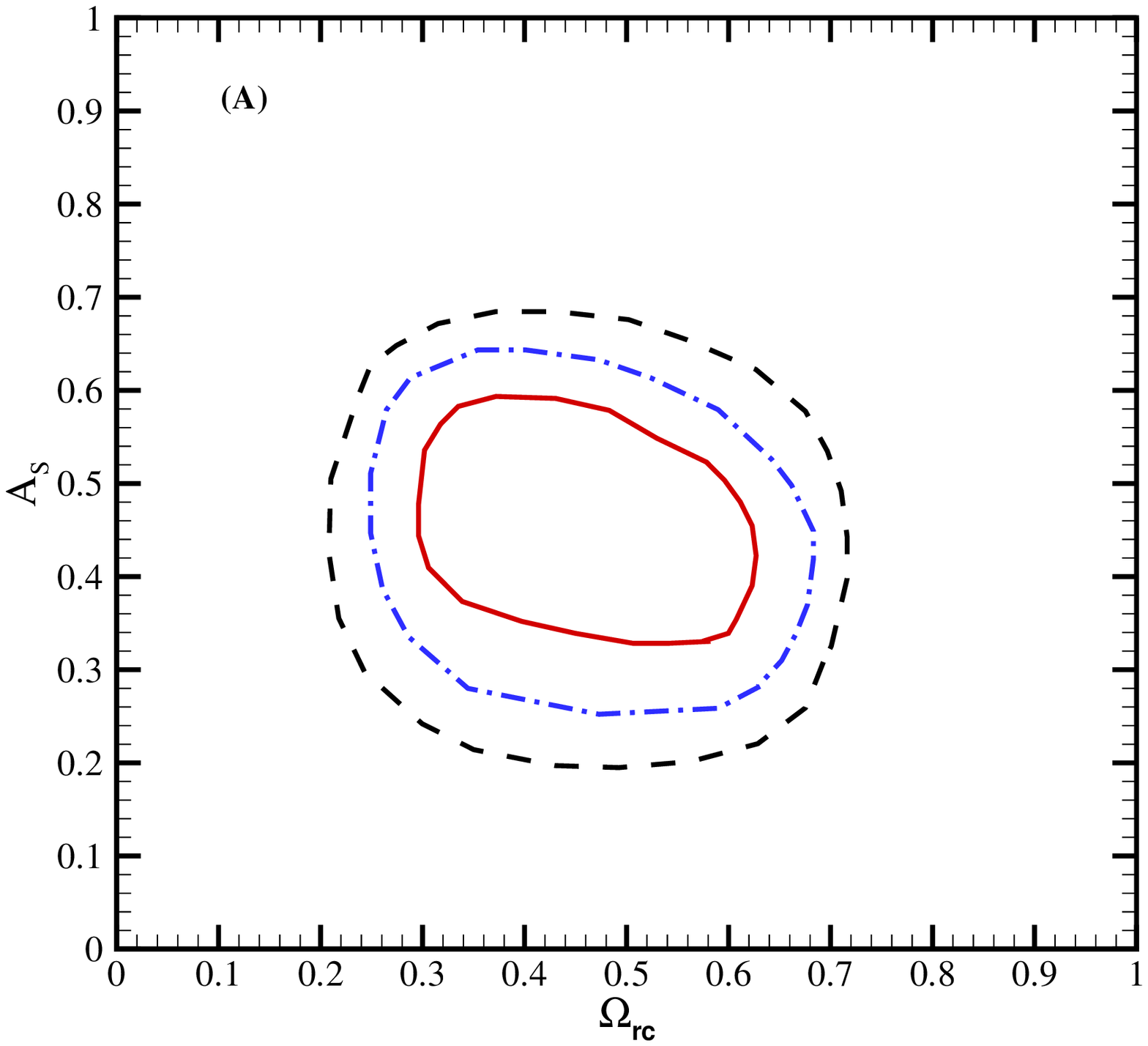} \vspace{2cm}\includegraphics{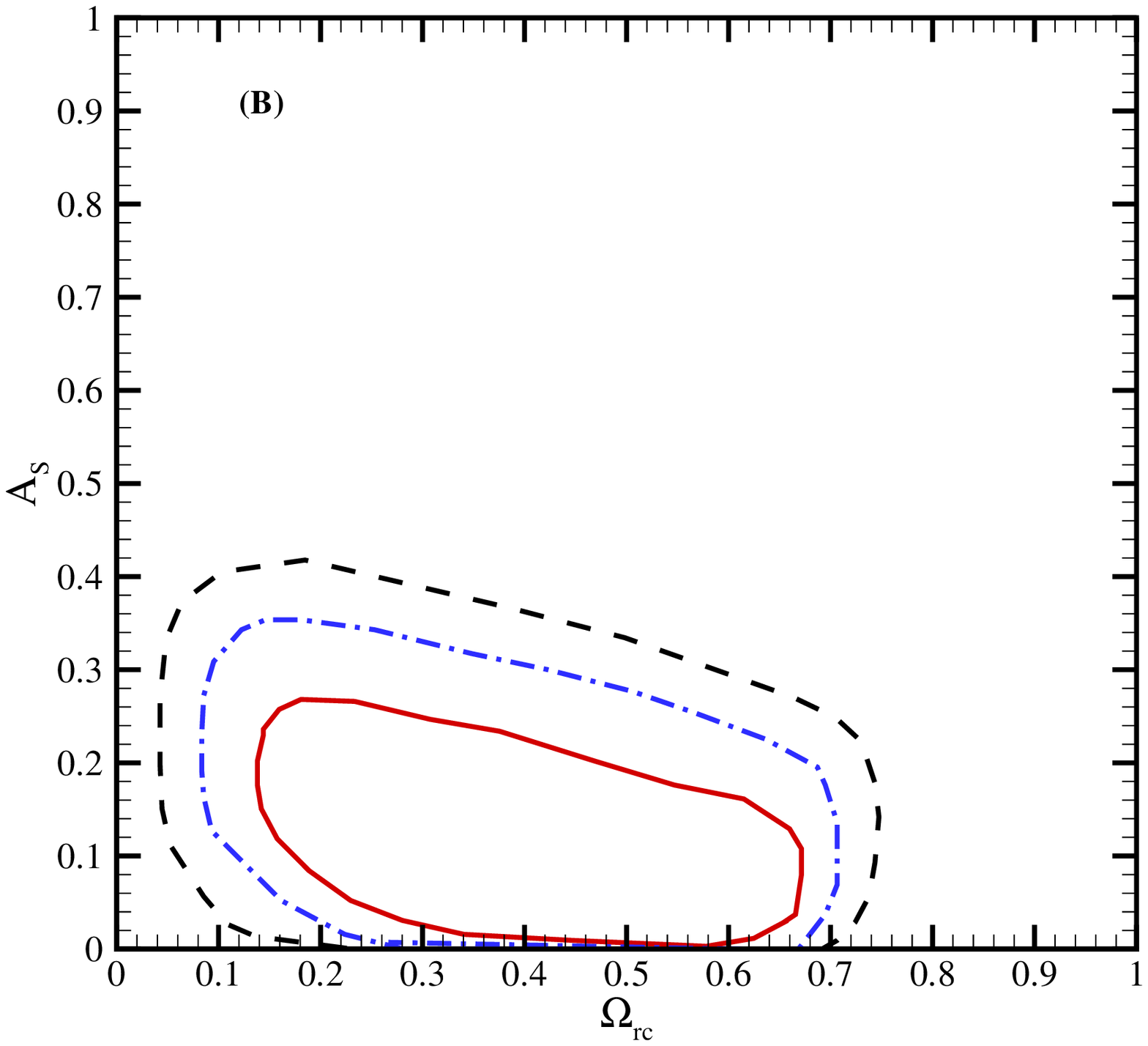}\vspace{3cm}
\end{center}
\end{figure}

\begin{figure}[htp]
\begin{center}\includegraphics{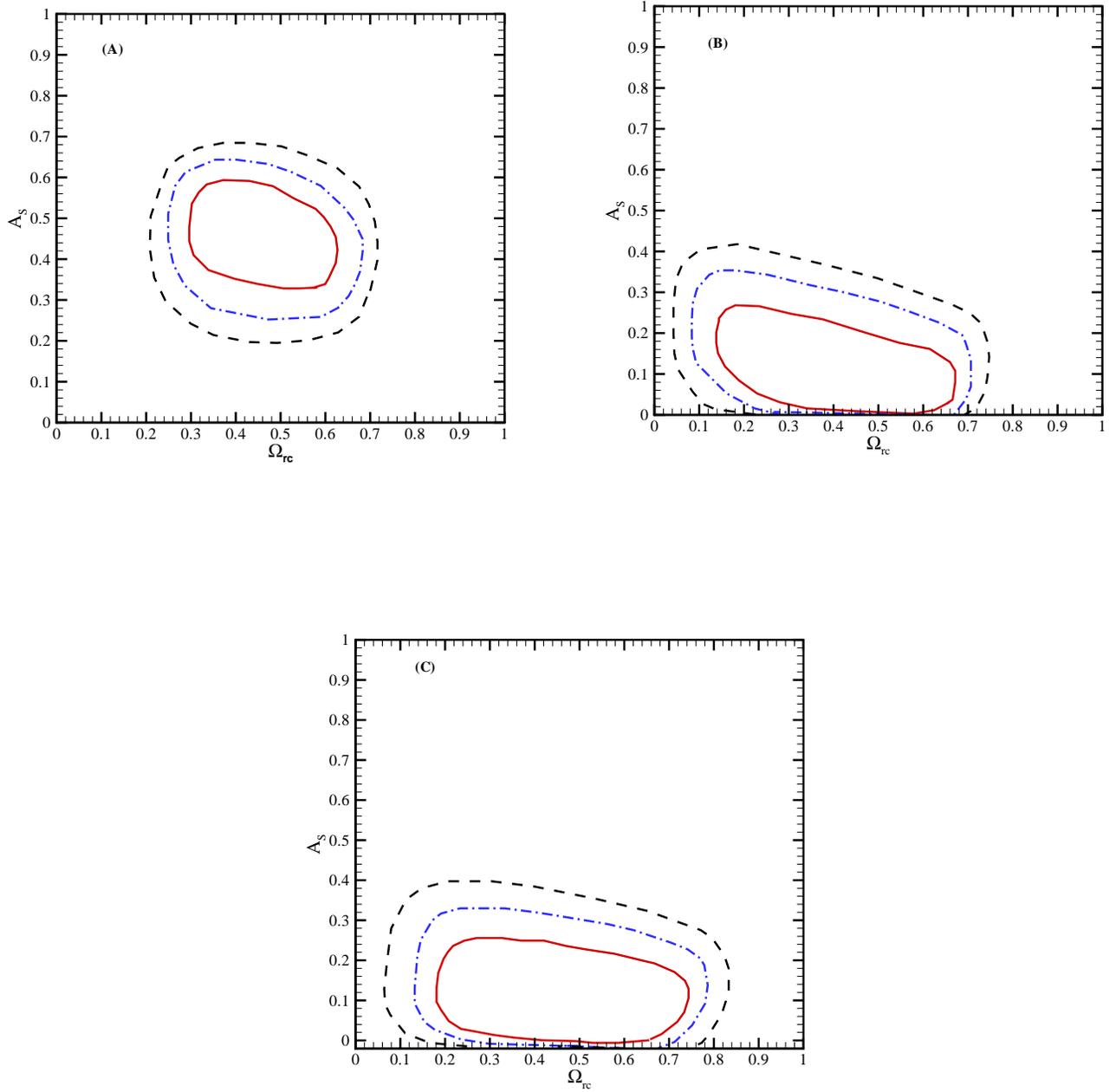} \vspace{4cm}
\end{center}
 \caption{\small {Confidence interval of\,
$\Omega_{{r}_{c}}$ and $A_{S}$ in $1\sigma$,\, $2\sigma$ and
$3\sigma$ level of confidence for the SNIa gold data (A), SNIa+CMB
joint data (B) and SNIa+CMB+SDSS joint data (C).}}
\end{figure}

\newpage

\begin{figure}[htp]
\begin{center}
\includegraphics{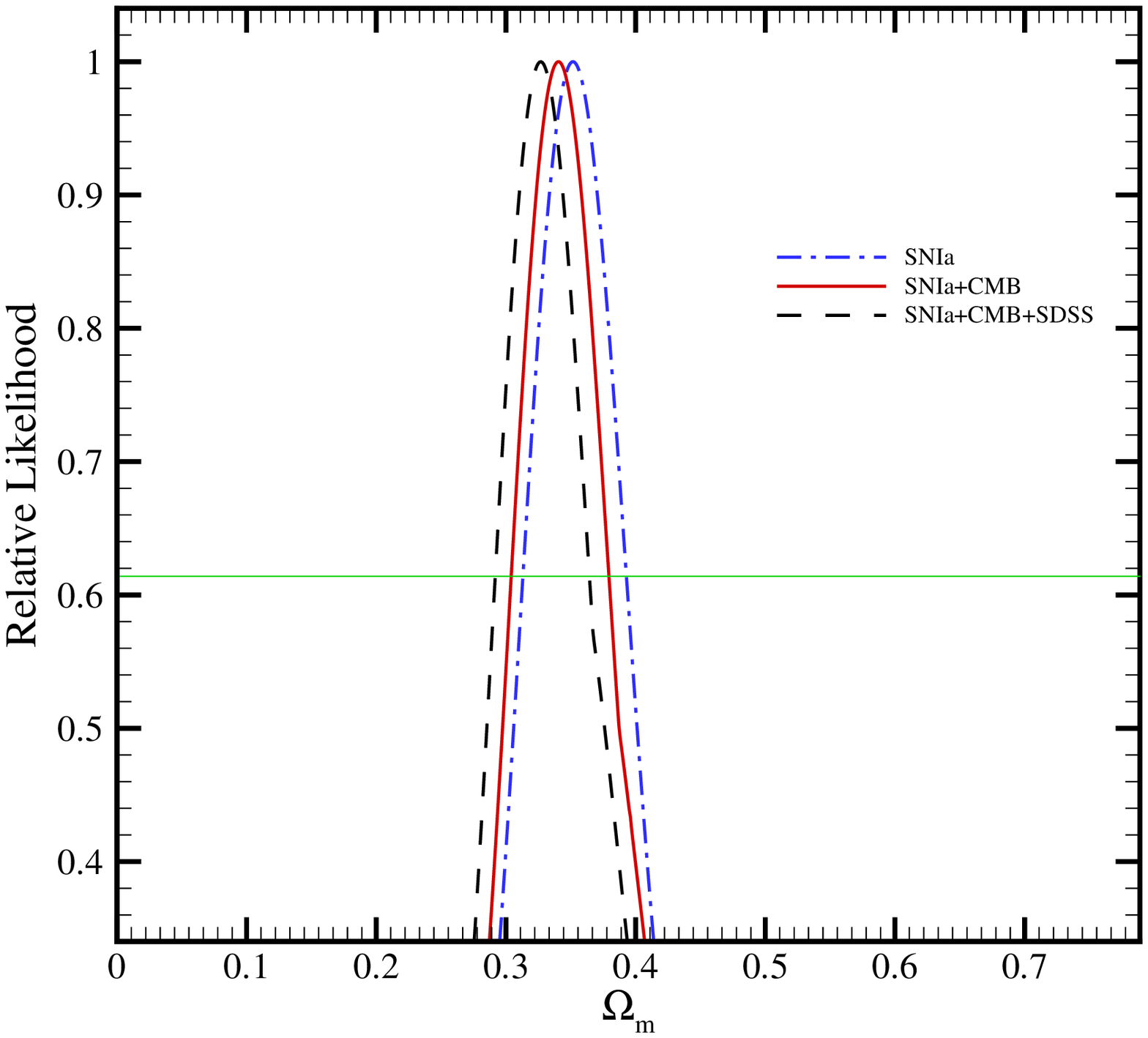} \vspace{2cm}\includegraphics{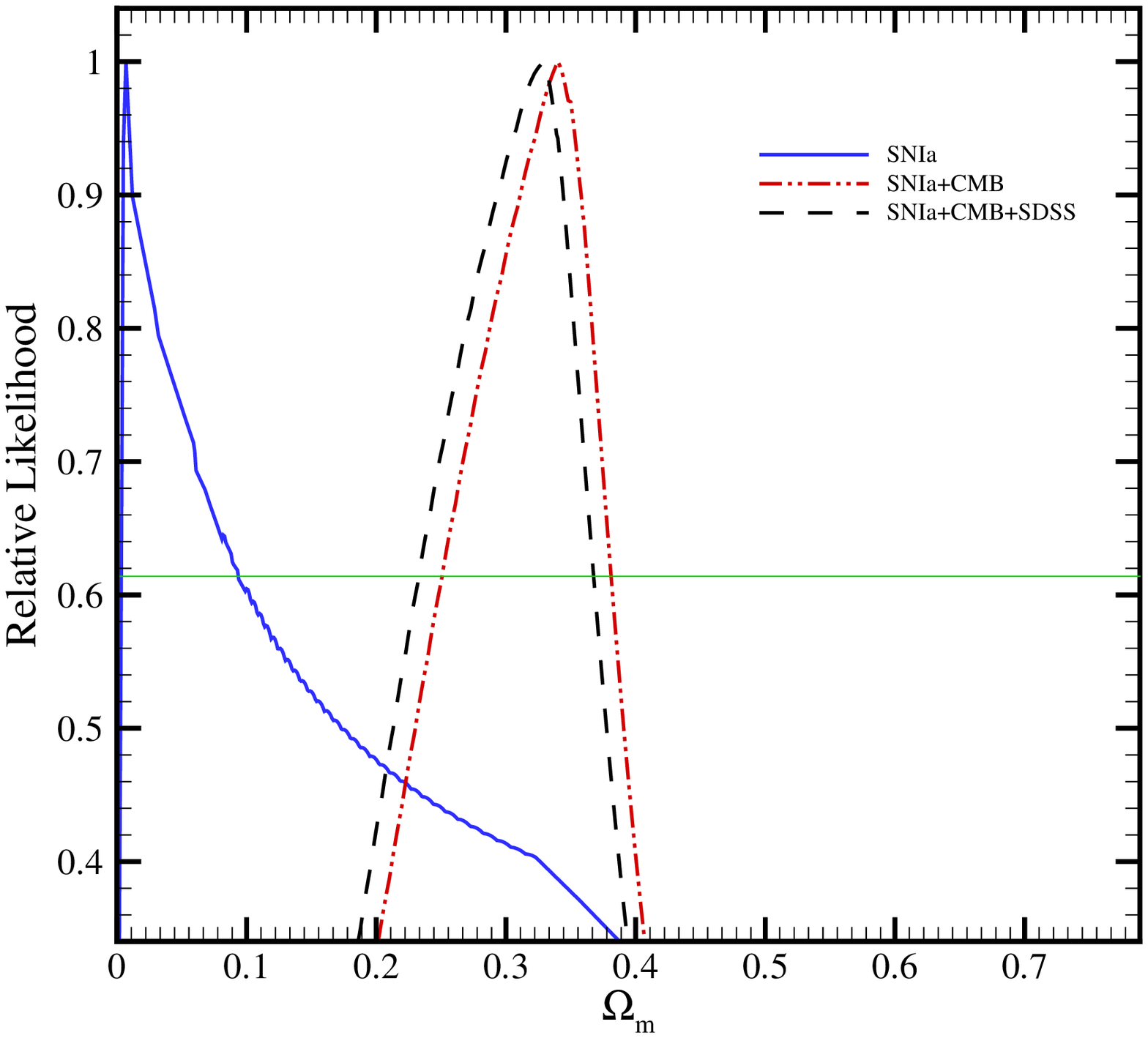}\vspace{5cm}
 \caption{\small {Marginalized relative likelihood with respect to parameter $\Omega_{m}$ fitted with SNIa
gold sample, SNIa+CMB and SNIa+CMB+SDSS experiments for $\Lambda$CDM
(left) and $\Lambda$DGP (right) respectively. The intersection of
the curve with the horizontal solid line is corresponding to the
bound with $1\sigma$ level of confidence. }}
\end{center}
\end{figure}

\section{Summary and Conclusion}

In this paper, we have considered a braneworld induced gravity
scenario which contains curvature effect of the Gauss-Bonnet term in
the bulk and a Chaplygin gas component in the brane action. We have
investigated the cosmological dynamics and the late-time behavior on
the brane. We have shown the possibility of getting accelerated
expansion in this scenario. We have studied the effective
phantom-like behavior on the brane. This behavior can be realized in
this scenario and the universe enters into the phantom phase at
$z\approx0.12$. We note that this effective behavior could be
deduced also in some braneworld models (such as $\Lambda$DGP and
pure GBIG scenario) with number of parameters less than our proposed
model. However, in such models the effective phantom picture breaks
down, and the effective equation of state parameter blows up at a
redshift in the past. In our model, inclusion of the Chaplygin gas
fluid on the brane saves the phantom-like prescription for a wide
range of redshifts. In fact, there is no break down of the effective
phantom picture in the presence of the Chaplygin component on the
brane. From another perspective, existence of the chaplygin gas
component on the brane, leads to a smooth crossing of the phantom
divide line by the effective equation of state parameter of the
model. We have studied also the notion of dynamical screening of the
brane cosmological constant in this generalized setup. Finally, We
have confronted the model with observational data from type Ia
Supernovae, Cosmic Microwave Background and Baryon Acoustic
Oscillations to constraint the model parameter space. The results of
comparison of the model parameters space with the observational data
are summarized in tables 1 and 2, and corresponding figures 8-10. We
have shown that up to the analysis on the parameters space performed
here ( table 2 and 3), this model has better agreement with
observations than the $\Lambda$DGP and $\Lambda$CDM. In other words,
existence of a Chaplygin component on the DGP brane with curvature
effect,  brings the scenario to be more viable in observational
viewpoint than the pure $\Lambda$DGP case. This result is due to the
wider parameter space and also further degrees of freedom accessible
in the model discussed here. \\

{\bf Acknowledgement} \\
We would like to thank Dr M. Sadegh Movahed for his invaluable
contribution in this work. The work of KN is supported partially by
the Research Institute of Astronomy and Astrophysics of Maragha,
IRAN.

\end{document}